\documentclass[aps,prb,amsmath,amssymb,twocolumn,superscriptaddress,floatfix]{revtex4-1}
\usepackage{graphicx}
\usepackage{hyperref}

\begin{document}
	
	\title{Magnetic and Electronic Properties of Single-Crystalline BaCoSO}
	
	\author{Yuanhe Song}
	\author{Xianshi Liu}
	\affiliation{State Key Laboratory of Surface Physics, Department of Physics, and Advanced Materials Laboratory, Fudan University, Shanghai 200438, China}
	
	\author{Dong Wu}
	\affiliation{International Center for Quantum Materials, School of Physics, Peking University, Beijing 100871, China}
	
	\author{Qi Yao}
	\author{Chenhaoping Wen}
	\affiliation{State Key Laboratory of Surface Physics, Department of Physics, and Advanced Materials Laboratory, Fudan University, Shanghai 200438, China}
	
	\author{Maxim Avdeev}
	\email{max@ansto.gov.au}
	\affiliation{Australian Nuclear Science and Technology Organisation, Lucas Heights, NSW 2234, Australia}
	
	\author{Rui Peng}
	\author{Haichao Xu}
	\author{Jin Miao}
	\author{Xia Lou}
	\author{Yifei Fang}
	\author{Binglin Pan}
	\affiliation{State Key Laboratory of Surface Physics, Department of Physics, and Advanced Materials Laboratory, Fudan University, Shanghai 200438, China}
	
	\author{Nanlin Wang}
	\affiliation{International Center for Quantum Materials, School of Physics, Peking University, Beijing 100871, China}
	\affiliation{Collaborative Innovation Center of Quantum Matter, Beijing 100871, China}
	
	\author{Darren C.\ Peets}
	\email{dpeets@nimte.ac.cn}
	\affiliation{State Key Laboratory of Surface Physics, Department of Physics, and Advanced Materials Laboratory, Fudan University, Shanghai 200438, China}
	\affiliation{Ningbo Institute of Materials Technology and Engineering, Chinese Academy of Sciences, Ningbo, Zhejiang 315201, China}
	
	\author{Donglai Feng}
	\email{dlfeng@fudan.edu.cn}
	\affiliation{State Key Laboratory of Surface Physics, Department of Physics, and Advanced Materials Laboratory, Fudan University, Shanghai 200438, China}
	\affiliation{Hefei National Laboratory for Physical Science at Microscale and Department of Physics, University of Science and Technology of China, Hefei, Anhui 230026, China}
	\affiliation{Collaborative Innovation Center of Advanced Microstructures, Nanjing 210093, China}
	
%	\date{\today}
	
	\begin{abstract}	
	  Doped BaCoSO was recently predicted to be a high-temperature superconductor in a new class based on Co and Ni. Using a Co--S self flux method, we synthesized single crystals of the antiferromagnetic insulator BaCoSO. Our magnetic and specific heat measurements and neutron diffraction provide details of its magnetic anisotropy and order. Its band gap was determined to be about 1.3\,eV by our measurements of its photoemission spectrum and infrared optical conductivity. %Dosing potassium onto its cleaved surface to suppress the magnetism and induce conductivity shifted the bands by up to 2\,eV, but we were unable to observe states at the Fermi level.
          Our results can pave the way to exploring the predicted superconductivity in this Co-based material.
	\end{abstract}
	
	%%%%%%%%%%%%%%%%%%%%%%%%%%%%%%%%%%%%%%%%%%%%%%%%%%%%%%%%%%%%%%%%%%%%%%%%%%%%
	\maketitle
	%%%%%%%%%%%%%%%%%%%%%%%%%%%%%%%%%%%%%%%%%%%%%%%%%%%%%%%%%%%%%%%%%%%%%%%%%%%%
	
	\section{Introduction}
	
	The high-temperature cuprate\cite{Bednorz1986} and iron-based superconductors\cite{Kamihara2006,Kamihara2008} were discovered during the past decades, but the underlying mechanisms of their unconventional high-temperature (high-$T_c$) superconductivity are not fully understood. Possible rules and factors governing high-$T_c$ superconductivity can be extracted from the two known families, suggesting theoretical frameworks that may explain the observed physics. Recent theoretical work found that both families of high-$T_c$ superconductors share a special electronic property in that the crucial $3d$-orbitals have a band filling that maximizes the antiferromagnetic (AFM) superexchange coupling\cite{Hu2015,Hu2016b}, through a collaborative interaction between the Fermi surface and short-range magnetic interactions\cite{Hu2012,Hu2016}. This led directly to predictions of new classes of high-$T_c$ superconductors hosted in doped planes of Ni$^{3+}$ and Co$^{2+}$ in trigonal bipyramidal coordination\cite{Hu2015,Hu2017b}. To date, it has proven difficult to verify this bold, concrete prediction, due to the difficulties encountered in actually preparing the predicted compounds.
	
	Subsequent theoretical work has indicated that the same physics should also be manifested in tetrahedrally-coordinated Co$^{2+}$, and families of chalcogenides and oxychalcogenides, such as BaCo$A$O ($A$=S, Se) and ZnCoS$_2$, were suggested\cite{Le2017,Li2017}. Adjusting the carrier concentration, for instance through chemical doping or applied pressure, may induce superconductivity in these materials\cite{Le2017}. These predictions are more readily tested, in particular since the synthesis of barium thio-oxocobaltate (BaCoSO) powder was reported not long before the prediction\cite{Valldor2015}. This insulator is isostructural to BaZnSO, crystallizing in an orthorhombic lattice (space group $Cmcm$) with the Co$^{2+}$ tetrahedrally coordinated to two sulfur and two oxygen anions. Co atoms form a square lattice buckled along the $c$ direction, and CoSO$^{2-}$ layers stack alternately with Ba$^{2+}$ along the $b$ axis. BaCoSO was proposed to have an AFM ground state based on density-functional theory (DFT) calculations\cite{Valldor2015}, and this was verified by neutron powder diffraction\cite{Salter2016}. If this compound can be made to superconduct, for instance upon hole or electron doping or the application of high pressure, it would constitute a class of high-$T_c$ superconductors discovered based on theoretical predictions, mark a  leap toward an understanding of the existing two classes of high-$T_c$ materials, and pave the way toward attaining higher transition temperatures by design.
	
	In this paper, we report the growth of high-quality single crystals of undoped BaCoSO. To clarify its magnetic ground state, we report its magnetic properties as a function of crystallographic orientation, and reexamine its magnetic order through neutron diffraction, resulting a magnetic space group different from that reported previously\cite{Salter2016}. We also report its band gap based on angle-resolved photoemission spectroscopy (ARPES) and optical spectroscopy measurements. %Dosing K atoms onto the surface was able to shift the valence bands by $\sim$\,eV, implying that doping of electrons into a surface layer was achieved, but charging effects prevented direct observation of the resulting bands.
	
	\section{Experimental}
	
	\subparagraph{Sample preparation}
	
	\begin{figure}[htb]
		\includegraphics[width=\columnwidth]{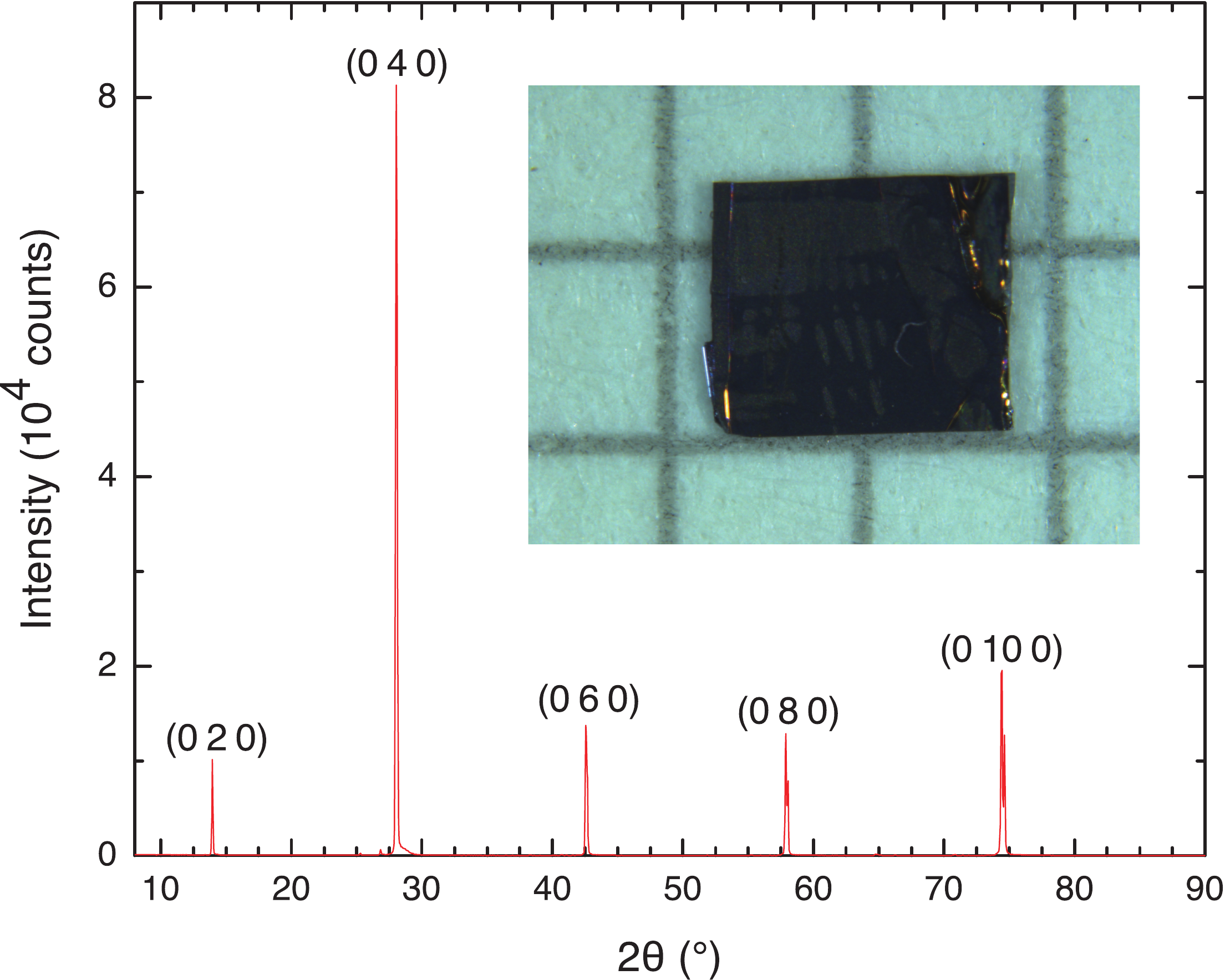}
		\caption{X-ray diffractogram and optical micrograph of the $b$-axis face of a BaCoSO single crystal. The single crystal is on a millimeter grid.}
		\label{fig:Xray}
	\end{figure} 
	
	Single-crystalline BaCoSO was obtained using a self-flux method. Following Ref.~\onlinecite{Salter2016}, we prepared precursor BaO by heating BaCO$_3$ powder (Alfa Aesar, 99.99\%) at 1090\,{$^\circ$C} for 12\,h under dynamic vacuum at $\sim$10$^{-5}$\,mbar. The BaO was then mixed with Co powder (Alfa Aesar, 99.998\%) and S granules (Aladdin, 99.999\%) under argon in the molar ratio 1:6:4.20. The ratio of excess Co to S corresponds to the eutectic at Co:S = 61:39 and 875\,$^\circ$C\cite{Sharma1979}, which serves as a flux with ratio 5:1 based on Co. The choice of a self flux eliminates the possibility of contamination by the components of the flux. The ground powder was packed into an alumina crucible, sealed inside a quartz tube and loaded into a crucible furnace. The charge was heated up to 1000\,$^\circ$C, held there for 4 hours, then slowly cooled to 880\,$^\circ$C at $-1.25\,^\circ$C/h. After several hours at this temperature, the quartz tube was removed to a centrifuge and the flux decanted through quartz wool, isolating the BaCoSO crystals. Crystals took the form of thin rectangular platelets, of size up to $2\times 2\times 0.2$\,mm$^3$; an example is shown in the Fig.~\ref{fig:Xray} inset. By an electron probe micro-analyzer (Shimadzu EPMA-1720), the average composition of our single crystals was determined to be BaCo$_{0.96(1)}$S$_{0.95(2)}$O$_{1.6(2)}$ (normalized to Ba), where the ratio of the first three elements is nearly consistent with the ideal stoichiometry. The high oxygen content is attributed to surface contamination, most likely adsorbed water and CO$_2$ from the air.
	
	\subparagraph{Diffraction}
	
	\begin{figure}[htb]
		\includegraphics[width=\columnwidth]{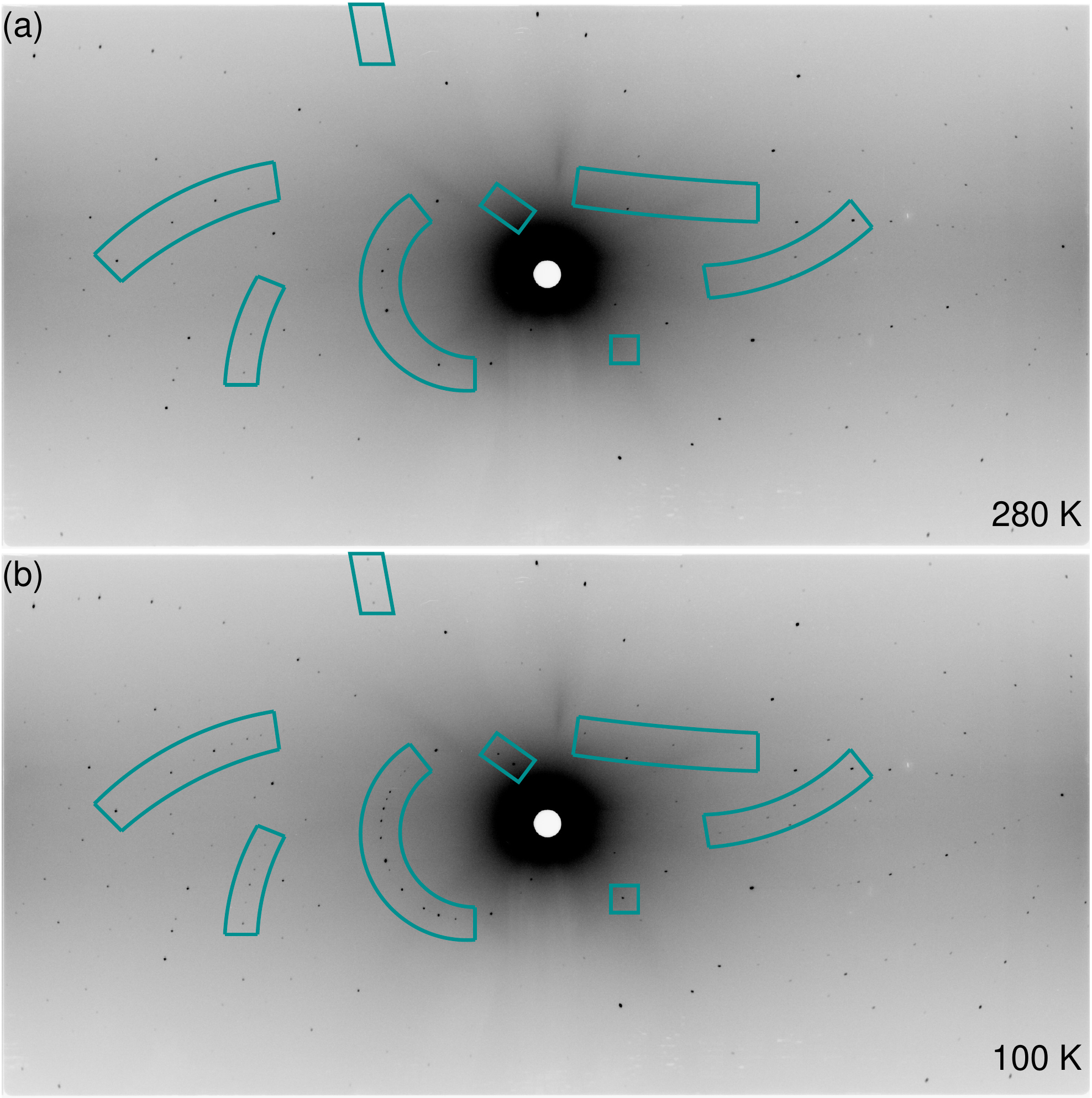}
		\caption{\label{fig:spots}Neutron Laue patterns of BaCoSO at (a) 280 and (b) 100\,K. Boxes highlight several regions where the patterns differ due to the presence of magnetic Bragg reflections at low temperature.}
	\end{figure}
	
	\begin{figure*}[htb]
		\includegraphics[width=\textwidth]{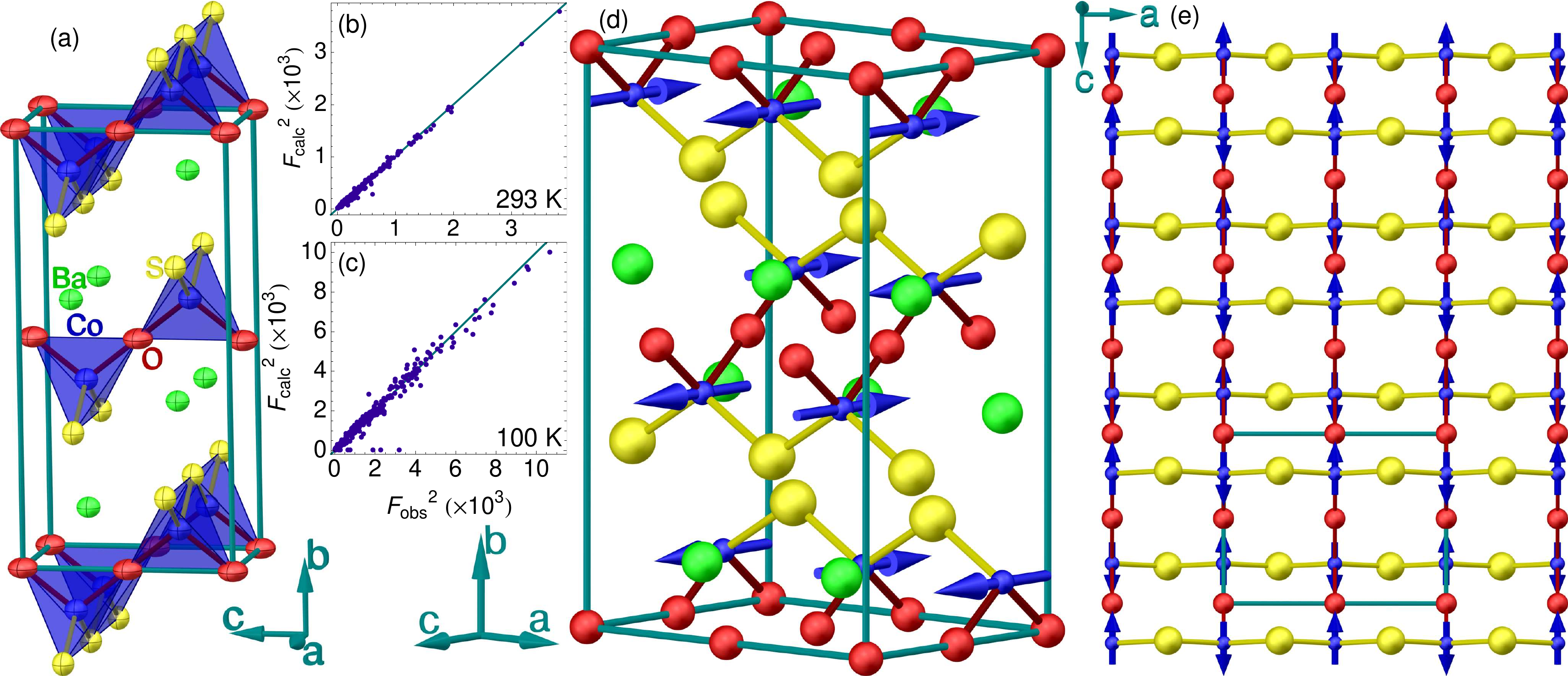}
		\caption{\label{fig:M1}(a) Refined crystal structure of BaCoSO at room temperature (space group $Cmcm$, \#63). Green, red, and yellow indicate barium, oxygen, and sulphur, respectively, while the blue tetrahedra indicate the coordination environment of Co. Atoms are shown as 99\%\ probability ellipsoids. (b) Comparison of structure factors $F_{calc}^2$ vs.\ $F_{obs}^2$ for the room temperature refinement, indicating the quality of the fit. (c) $F_{calc}^2$ vs.\ $F_{obs}^2$ for the 100\,K refinement, including magnetic peaks. Low-intensity peaks ($<$3 standard deviations above background) were not refined and are not included here. (d) Refined magnetic structure of BaCoSO at 100\,K (Shubnikov group $P_abcm$, Belov-Neronova-Smirnova \#57.386). (e) View of the CoSO$^{2-}$ plane at $y=0$, showing the in-plane antiferromagnetic spin orientations.}
	\end{figure*}
	
	We performed X-ray diffraction on the large flat face of a crystal at ambient temperature in a laboratory diffractometer (Bruker D8 Discover) using Cu-K$\alpha$ radiation. 
	
	A single crystal was measured at room temperature, 280, and 100\,K using the KOALA white-beam neutron Laue diffractometer at the OPAL Research Reactor at ANSTO, in Australia\cite{KOALA2011}. Image data processing, including indexing, intensity integration, and wavelength distribution normalization were performed using LaueG\cite{LaueG}. Crystal and magnetic structure refinement was carried out using Jana2006\cite{Jana2006}.
	
	\subparagraph{Magnetic susceptibility}
	
	\begin{figure}[htb]
		\includegraphics[width=\columnwidth]{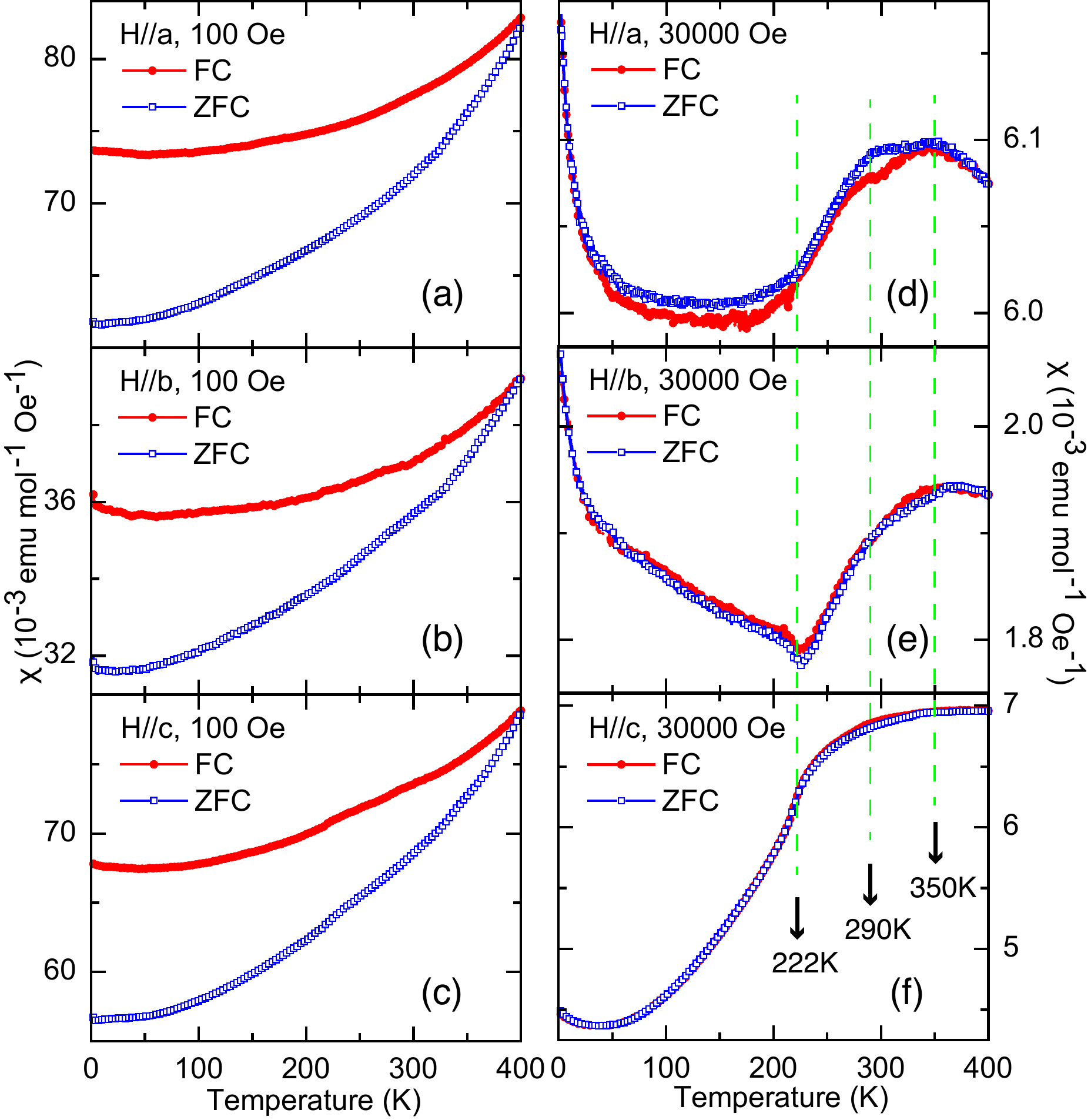}
		\caption{Magnetic susceptibility ($\chi$, $M/H$) {\it vs}.\ temperature. The left three panels were measured in a 100\,Oe applied field and the right in a 30000\,Oe field. The upper, middle and lower panels correspond to fields applied along $a$, $b$ and $c$, respectively. ZFC and FC magnetic susceptibility results are given by open squares and solid dots, respectively. The green dashed lines along $T$ = 222, 290 and 350\,K serve as guides to the eye.}
		\label{fig:Mag}
	\end{figure}
	
	DC magnetic susceptibility measurements were performed on single-crystalline BaCoSO in a magnetic property measurement system (MPMS SQUID-VSM, Quantum Design), for fields along its $a$, $b$ and $c$ axes. The crystal orientation was determined by x-ray Laue diffraction. During the MPMS measurement, the samples were first cooled down to 2\,K in zero magnetic field, then a field of 100 or 30000\,Oe was applied. The zero-field-cooling (ZFC) curves were measured on heating up to 400\,K, then field-cooling (FC) curves were obtained upon cooling back to 2\,K. 
	
	Although most of the Co--S flux was removed by centrifugation, the obtained crystals sometimes had flux remaining on their surface (see golden areas in Fig.~\ref{fig:Xray}). Such samples were polished using abrasive paper before magnetic measurements to reduce the influence from impurities. 
	
	\subparagraph{Specific Heat}
	
	\begin{figure}[htb]
		\includegraphics[width=\columnwidth]{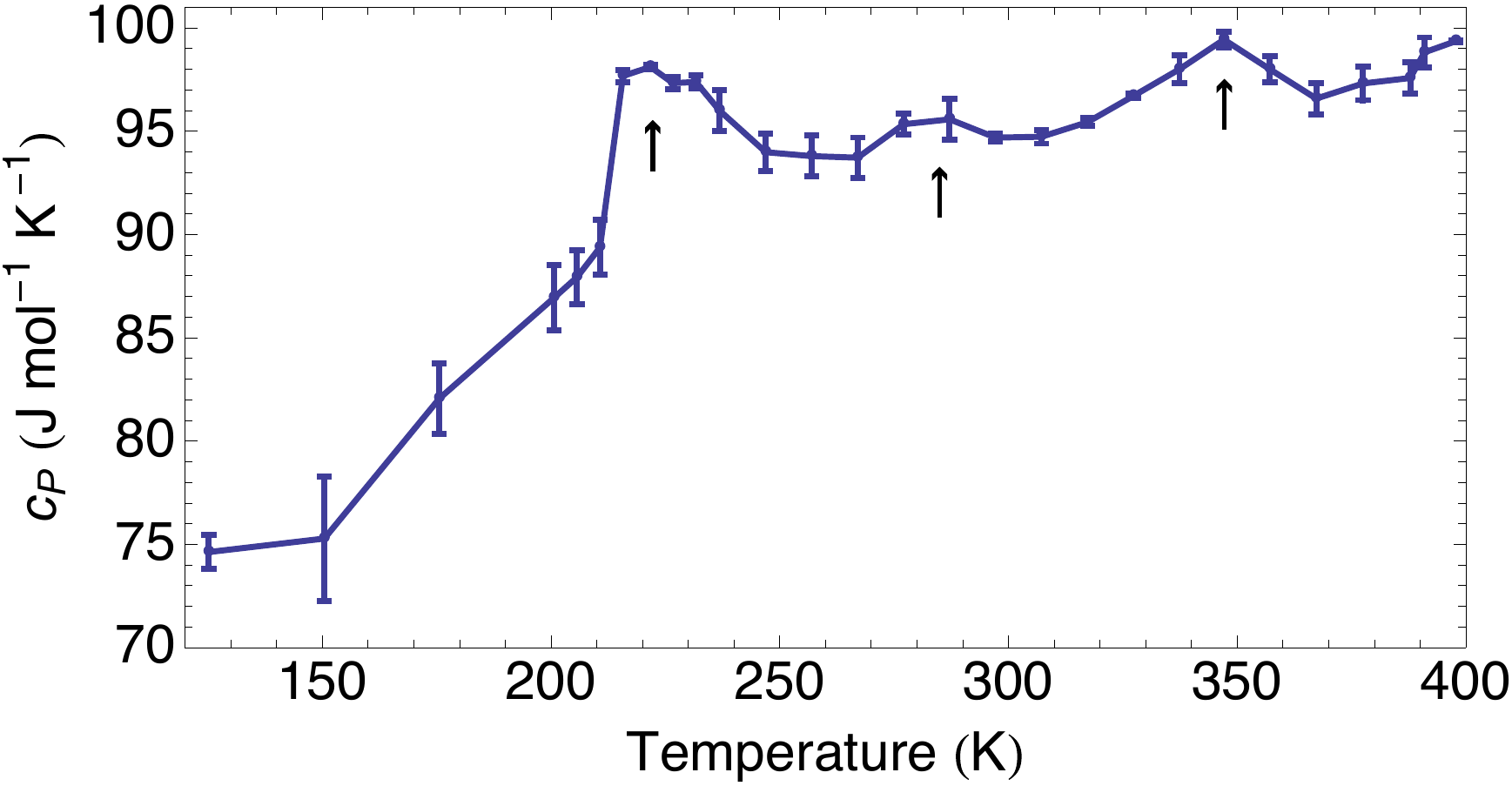}
		\caption{Specific heat of BaCoSO at high temperatures. Multiple weak peaks visible in this temperature range are pointed out by arrows.}
		\label{fig:cP}
	\end{figure}

	Specific heat was measured on cooling from 400 to 125\,K on a mosaic of three crystals using a physical properties measurement system (PPMS, Quantum Design) by the relaxation time method. The crystals were attached to the sample stage using Apiezon H-grease to avoid the N-grease glass transition encountered in this temperature range. At least three measurements were taken at each temperature; due to evidence that the sample did not equilibrate rapidly, the first measurement at each temperature was discarded as a precaution before averaging the rest.

	\subparagraph{Photoemission and optical study}
	
	\begin{figure*}[htb]
		\includegraphics[width=0.9\textwidth]{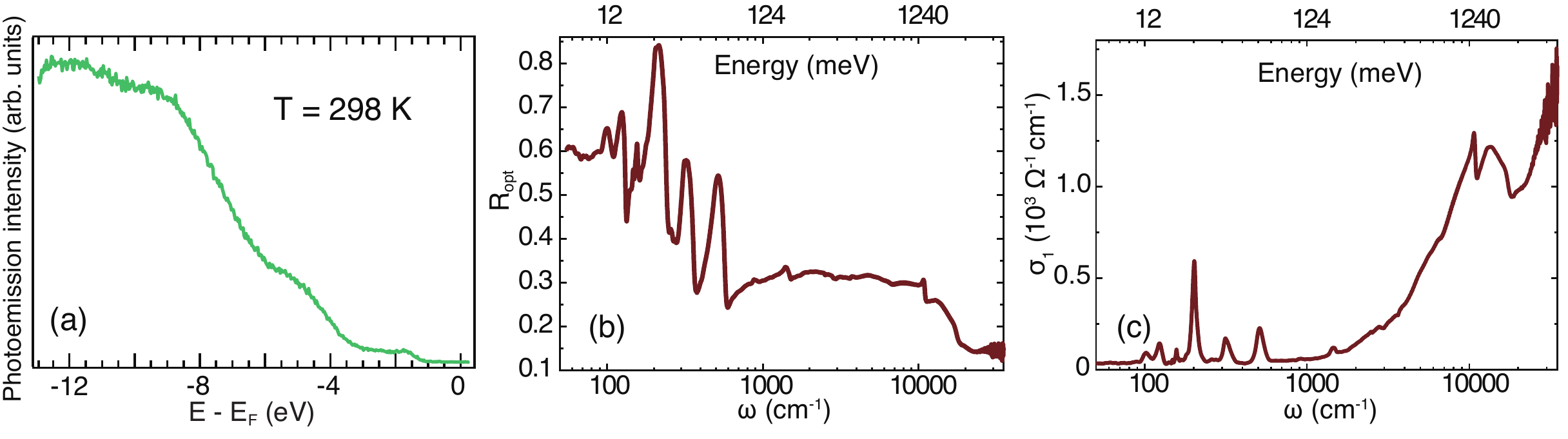}
		\caption{Photoemission and optical study of BaCoSO. (a) Angle-integrated photoemission spectrum at room temperature ($T$ = 298\,K); (b) optical reflectivity $R_{opt}$ and (c) optical conductivity ${\sigma}_1$ as a function of frequency $\omega$ at room temperature.}
		\label{fig:spectrum}
	\end{figure*}
	
%	\begin{figure}[htb]
%		\includegraphics[width=0.5\columnwidth]{K_dosing.pdf}
%		\caption{Photoemission spectra taken at 30\,K with different K doping levels, which increased from \#1 to \#7. The short vertical dashes under the valence band peak around $-$9\,eV show its shift induced by K dosing. The red dashed line serves as a guide to the eye. The time and current in the legend correspond to the additional K-dosing after the preceding dataset.}
%		\label{fig:K_dosing}
%	\end{figure}
	
	Photoemission measurements were performed on an in-house ARPES system with a Fermi Instruments 21.2\,eV helium discharge lamp, using a VG-Scienta DA30 electron analyzer. The energy resolution is 6\,meV and angular resolution is 0.3\,$^\circ$. The sample was cleaved under ultrahigh vacuum. 
	
%	In the K-dosing ARPES study, after the sample was cleaved at 30\,K, an SAES K dispenser was used to gradually dose potassium onto the surface. The K coverage was increased between successive scans in the measurements. The deposition time and heating current of the K dosing before each measurement were gradually increased to enhance the dosing effect.
	
	The optical reflectance measurements were performed on Bruker 80v spectrometer in the frequency range from 55 to 35000\,cm$^{-1}$. An {\sl in situ} gold and aluminum overcoating technique was used to obtain the reflectivity $R(\omega)$. The real part of the conductivity $\sigma_1(\omega)$ is obtained by Kramers-Kronig transformation of $R(\omega)$. A constant relation was used for low-frequency extrapolation; on the high-frequency side a ${\omega}^{-1}$ relation was used up to 300000\,cm$^{-1}$, above which ${\omega}^{-4}$ was applied.

	\section{Results}
	
	\subsection{Crystal and Magnetic Structure}
	
	The X-ray diffraction result (Fig.~\ref{fig:Xray}) shows a series of sharp peaks corresponding to the crystallographic $b$ axis, with lattice parameter $b$ = 12.75(9)\,\AA, consistent with the published refinements\cite{Valldor2015,Salter2016}.
	
	In our neutron result, sharp Laue diffracted spots indicate excellent crystal quality [see Fig.~\ref{fig:spots}(a)]. Systematic absences were found consistent with the $Cmcm$ space group, in agreement with the previous x-ray powder diffraction crystal structure determination\cite{Valldor2015}. The data collected at 100\,K clearly show additional diffraction peaks indicating long-range magnetic order [see Fig.~\ref{fig:spots}(b)]. 
	
	Analysis of the room temperature (293\,K) neutron Laue diffraction data was performed using the structure reported in Ref.~\onlinecite{Valldor2015} as the starting model (space group $Cmcm$, \#~63), to verify the structure and atomic positions. The results of the crystal structure refinement are presented in Tabs.~\ref{tab:NXD300a}, \ref{tab:NXD300b}, and \ref{tab:NXD300c} and the crystal structure is shown in Fig.~\ref{fig:M1}(a). The atomic positions essentially agree with previous work\cite{Valldor2015}.
	
	The additional peaks in the low-temperature Laue diffraction data indicating long-range magnetic ordering could be indexed by a unit cell with doubled $a$ lattice parameter, i.e. propagation $k$-vector (1/2,0,0), in agreement with Ref.~\onlinecite{Salter2016}. Exhaustive testing of the models compatible with $k$=(1/2,0,0) and the symmetry of the $4c$ Co site of the $Cmcm$ space group unambiguously pointed to the magnetic structure shown in Fig.~\ref{fig:M1}(d) (magnetic space group $P_abcm$, Belov-Neronova-Smirnova \#57.386). This antiferromagnetic structure with Co moments along the $c$ axis is essentially the same as that described by Salter et al.\cite{Salter2016}, who however used a lower symmetry to describe it. The magnetic space group $P_cma2$ reported in Ref.~\onlinecite{Salter2016} (Belov-Neronova-Smirnova \#28.94) is a subgroup of $P_abcm$ and splits the original single Co site into two sites with independent magnetic moments. However, once the moments of the two Co sites of $P_cma2$ are constrained to be equal, as was also done in Ref.~\onlinecite{Salter2016} (Table S12), the structure corresponds to $P_abcm$ with a single Co site. To verify the correct assignment of the magnetic space group, we also carried out the analysis using $P_cma2$ with two independent Co sites and confirmed that this does not result in any further improvement of the refinement quality. Thus, we conclude that the correct magnetic symmetry of BaCoSO is $P_abcm$ with a single Co magnetic site and the magnetic transition occurs via a single irreducible representation $mSM1$, precluding spin components 
	along $a$ or $b$. The refined value of the Co magnetic moment, 3.017(15)\,$\mu_B$ (compare 2.75(2)\,$\mu_B$ reported in Ref.~\onlinecite{Salter2016}), is practically equal to the theoretical spin-only value of 3$\mu_B$ expected for high-spin $3d^7$ Co$^{2+}$.
	
	The refined magnetic structure is presented in Tab.~\ref{tab:NXD100b} and Figs.~\ref{fig:M1}(d) and \ref{fig:M1}(e), with refinement details in Tab.~\ref{tab:NXD300a}. Plots of $F_{calc}^2$ vs.\ $F_{obs}^2$ for the room temperature and 100\,K refinements are presented in Figs.~\ref{fig:M1}(b) and \ref{fig:M1}(c), respectively, and demonstrate the quality of the refinements.
	
	\subsection{Magnetic Anisotropy and Order}
	
	The magnetic susceptibility results are shown in Fig.~\ref{fig:Mag} for fields of 100 and 30000\,Oe. Since the low field susceptibility decreases on cooling throughout the temperature range measured, we can conclude that the material is antiferromagnetic with significant magnetic correlations up to at least 400\,K. Higher-field data show a clearer N\'eel temperature $T_N \simeq 222$\,K, consistent with previous work on powder{\cite{Valldor2015}}. 
	
	However, single crystals also enable a field-direction-dependent study --- susceptibility is shown in Fig.~\ref{fig:Mag} for fields along the three different crystalline axes. The susceptibility values for fields along the $a$ and $c$ directions are larger than for $b$ in both fields. Above 50\,K, a strongly anisotropic temperature dependence of the magnetic susceptibility is seen in the higher-field data, particularly in the scale of the changes. The susceptibility varies several times more for fields along $c$ than along $a$ or $b$. The susceptibility measured in powder samples under similar high fields\cite{Valldor2015} should thus be dominated by the $c$ axis. ZFC and FC data show a clear difference in low field while they are nearly identical in high field. Besides the significant transition at 222\,K, some extra weak features are also visible around 290 and 350\,K, especially for the high field along $a$. Below 50\,K, low-temperature upturns are seen in high field for all field orientations with minimal anisotropy, which is suggestive of a minor ferromagnetic impurity phase, perhaps flux stuck to the edges of the crystal. This effect was also visible in a 3\,T field in the earlier powder work\cite{Valldor2015}. 
	
	The specific heat is shown in Fig.~\ref{fig:cP}. A series of weak peaks are observed around 220, 290, and 350\,K, corresponding to the features seen in the susceptibility. Specific heat probes bulk thermodynamic transitions, so these features would normally be attributed to magnetic transitions of BaCoSO. However, 290\,K corresponds to the antiferromagnetic transition of CoO at 289\,K\cite{Bizette1951,La1951,Herrmann1978}, and this peak is very weak, with an entropy release of only 0.07\,J/molK, so we attribute this feature to residual flux on the crystal edges. 350\,K is in a temperature range that does not have magnetic Bragg peaks, so this cannot be a transition into bulk long-range order, and it is also most likely extrinsic. Even if we assume that both transitions are intrinsic to BaCoSO the magnetic entropy released in this entire temperature range, as has been previously noted\cite{Valldor2015}, is extremely small. We estimate the entropy release at these two transitions as 0.80 and 0.19\,J/molK, respectively, totalling roughly 8.5\,\%\ of the entropy release of 11.53\,J/molK expected for $s=3/2$.

	\subsection{Photoemission and optical spectrum}
	
	Fig.~\ref{fig:spectrum}(a) presents the valence band of BaCoSO via the angle-integrated photoemission spectrum within 13\,eV of the Fermi level ($E_F$) at room temperature (298\,K). No Fermi drop is observed near the Fermi level. Some angle dependence appears in the angle-integrated spectra taken at different polar emission angles, but no clear energy bands with dispersive features were seen in the corresponding angle-resolved spectra (see the Supplemental Material for details\cite{SuppMat}). A peak near $E_F$ is observed at -1.77\,eV with a leading edge at $-$1.09\,eV, suggesting a lower limit for the gap of BaCoSO to be $\sim$1.1\,eV.
        In fact, the photoemission experiments suffered from charging problems due to the strongly insulating nature of this material, particularly at low temperatures.  The results quoted here are based on measurements performed at room temperature, using only 1/30 of our usual incident light intensity, under which conditions the influence of charging is under control (below the level of 0.01\,eV).  
        (See the Supplemental Material for details \cite{SuppMat}). 
	
	Figs.~\ref{fig:spectrum}(b) and (c) show optical reflectivity and the real part of the optical conductivity as a function of frequency $\omega$, respectively. In the low frequency limit, the optical reflectivity is around 0.6 rather than 1.0 while the optical conductivity approaches zero, showing typical insulating behavior. The foremost significant peak in the optical conductivity around 10700\,cm$^{-1}$ (excluding the minor phonon peaks between 100 and 1000\,cm$^{-1}$) corresponds to a band gap of about 1.3\,eV. This gap value is consistent with our photoemission result and also comparable to the 2\,eV band gap reported in the DFT calculation\cite{Valldor2015}.
	
%	Fig.~\ref{fig:K_dosing} presents the results of our surface K-dosing ARPES study at 30\,K. The figure gives a series of angle-integrated spectra taken at different K dosing levels, where the valence bands (VBs) are visible. The electron doping level increased from \#1 to \#7. Spectrum \#1 is for the freshly cleaved surface, while \#7 is so heavily doped that its VB becomes weak and featureless, while a new structure appears at $-$3.8\,eV, presumably a K Auger signal\cite{Petersson1977}. The behavior of the peak around $-$9\,eV indicates a shift of the valence band by about 2\,eV after the longest deposition time. However, no states emerge near $E_F$. (See the Supplemental Material for more details \cite{SuppMat}).  

	\section{Discussion}
	BaCoSO has a layered crystal structure with an orthorhombic lattice, so anisotropic magnetic properties would be expected. Comparing its magnetic response along different crystal axes, we see significantly greater susceptibility values for fields within the $ac$ plane. The magnetic structure determined based on DFT calculations\cite{Valldor2015}, neutron powder diffraction\cite{Salter2016}, and our own work [Fig.~\ref{fig:M1}(d)] indicates in-plane spin orientations, so this anisotropy implies that the magnetic moments are more easily rotated within the $ac$ plane than toward the $b$ axis. 
	
	The susceptibility above 220\,K is not paramagnetic, rather it decreases on cooling through most the temperature range probed, and only a small fraction of the magnetic entropy is released within this temperature range. This indicates that magnetic correlations set in at much higher temperatures. Earlier magnetization data on powder samples\cite{Valldor2015} exhibited a broad hump that could indicate a transition as high as 700\,K. This is unlikely to be a bulk N\'eel transition ($T_N$), however. Since BaCoSO is a layered material, magnetic correlations would be expected to develop within the plane at temperatures far above the bulk magnetic transition, only condensing into three-dimensional order at a much lower temperature. White-beam neutron Laue diffraction is not sensitive to the diffuse scattering that would be expected from this precursor short-range order, so we are unable to confirm this through direct observation. 
	
	Superconducting order competes with spin and charge order in the high-$T_c$ superconductors\cite{Ghiringhelli2012,Chang2012,Wu2011,Santi2013,Hucker2014}, but may rely crucially on spin fluctuations from the nearby AFM state. We sought to test predictions of similar physics in BaCoSO, whereby suppressed AFM order would lead to a robust $d$-wave-like superconducting state\cite{Le2017}, through performing ARPES after dosing potassium onto the cleaved surface to introduce carriers, but %we were unable to observe any states emerging near the Fermi level, despite K dosing shifting the valence band by as much as 2\,eV.
        these results were not reliable due to severe charging effects at low temperature, where bulk BaCoSO is strongly insulating.

	Comparing the band gap of 1.3\,eV observed in the optical conductivity and the gap of 1.1\,eV in the photoemission spectrum, we can deduce that the band immediately above the Fermi level is approximately located at $E_F$ + 0.2\,eV. A band shift of $\sim$2\,eV by surface potassium dosing also suggests that electrons can probably be doped into this material, although charging effects make this value imprecise and prevent strong conclusions. %Our non-observation of new states near $E_F$ may be due to strong charging effects at low temperature, as the strongly-insulating bulk was unable to compensate for the ejected photoelectrons [See the Supplemental Material for details \cite{SuppMat}].
        The comparison of ARPES and optical gaps indicates that, for chemical doping to turn BaCoSO into a metal, electron doping should be more readily realized than hole doping, although we note that we attempted to chemically dope the crystals with La on the Ba site, Fe/Ni on the Co site, and F on the O site, but without success. Besides introducing carriers, applying high pressure to enhance inter-layer interactions may be another good option for making this proposed superconductor superconduct.

	\section{Conclusion}
	
	In summary, our growth of single-crystalline BaCoSO allowed us to investigate details of its magnetic and electronic properties. The crystal and magnetic structure are closely similar to previous reports on powder, but the magnetic space group is different from that previously reported, and we see the full Co$^{2+}$ moment. We find evidence for magnetic transitions around 220 and possibly 350\,K, but the vast majority of the spin entropy is not released at these transitions. The photoemission and optical study identified the band gap of this material as around 1.3\,eV, and will help focus future work on obtaining a metallic ground state. %Attempts to induce the theoretically-predicted high-$T_c$ superconductivity via surface K dosing shifted the valence band by up to 2\,eV but failed to introduce observable states near the Fermi level. This result does not exclude the possibility of superconductivity in doped BaCoSO because charging effects at low temperature cannot be neglected, but it suggests that
        Hints from surface potassium surface dosing suggest that electron doping may be able to turn the material into a metal.
        In addition, applying high pressure to bulk samples or growing strained films may offer opportunities to realize a conducting state and verify the prediction of superconductivity in BaCoSO.
	
	\begin{acknowledgments}
		
		We gratefully acknowledge helpful discussions with J.\ P.\ Hu and experimental support by X.\ P.\ Shen and E.\ J.\ Cheng. This work is supported by Science Challenge Project (Grant No.~TZ2016004), the National Key R\&D Program of the MOST of China (Grant No.~2016YFA0300203), and the National Natural Science Foundation of China (Project Nos.~11650110428, 11421404, U1532266, 11790312, and 11674367). DCP is supported by the Chinese Academy of Sciences through 2018PM0036.
		
		%This research used resources of the Advanced Light Source, which is a DOE Office of Science user facility under contract No.~DE-AC02-05CH11231.
		
	\end{acknowledgments}
	
	\bibliography{BaCoSO}

	\appendix
	
	\section{Neutron Diffraction}
	
	\begin{table*}[htb]
		\caption{\label{tab:NXD300a}Refinement details for BaCoSO at room temperature and 100\,K. Note that white-beam neutron Laue diffraction is not sensitive to absolute lattice parameters. The magnetic refinement was actually performed in a doubled unit cell, with constraints to ensure the original $Cmcm$ symmetry was still obeyed, which inflates the number of total reflections reported here.}
		\begin{tabular}{lll}\\ \hline\hline
			Formula & BaCoSO & BaCoSO (magnetic)\\
			Temperature & 293\,K & 100\,K\\
			Space group & $Cmcm$ (\#~63) & $P_abcm$ (\#57.386)\\
			$a$ & 3.979\,\AA & 3.986\,\AA\\
			$b$ & 12.744\,\AA & 12.73\,\AA\\
			$c$ & 6.107\,\AA & 6.096\,\AA\\
			$Z$ & 4 & 4\\
			$F(000)$ & 428 & 130\\
			$\theta$ range & 4.62 to 50.88$^\circ$ & 4.5 to 45.3$^\circ$\\
			Index ranges & 0$\leq$$h$$\leq$7, 0$\leq$$k$$\leq$27, 0$\leq$$l$$\leq$11 & 0$\leq$$h$$\leq$6.5, 0$\leq$$k$$\leq$25, 0$\leq$$l$$\leq$12\\
			Total reflections & 565 & 1967\\
			Reflections $I$$>$3$\sigma(I)$ & 352 & 404\\
			Goodness of fit, all data & 1.40 & 1.89\\
			Goodness of fit, $I$$>$3$\sigma(I)$ & 1.58 & 2.83\\
			$R$ factors, $I$$>$3$\sigma(I)$ & $R_1$=0.0363, $wR_2$=0.0290 & $R_1$=0.0745, $wR_2$=0.0597\\
			$R$ factors, all data & $R_1$=0.1184, $wR_2$=0.0327 & ---\\
			Extinction coefficient & 490(50) & 310(80)\\ \hline
		\end{tabular}
	\end{table*}
	
	%% Doubling the unit cell with constrained positions creates a large number of reflections with zero intensity, making $R$ factors on all data meaningless
	
	\begin{table}[htb]
		\caption{\label{tab:NXD300b}Refined atomic positions for BaCoSO at room temperature.}
		\begin{tabular}{lccr@{.}lr@{.}lr@{.}l}\\ \hline
			Site & Mult.& $x$ & \multicolumn{2}{c}{$y$} & \multicolumn{2}{c}{$z$} & \multicolumn{2}{c}{$U_{eq}$}\\ \hline \hline
			Ba & $4c$ & 0 & 0&38622(4) & 0&25 & 0&00886(13)\\
			Co & $4c$ & 0 & 0&09449(7) & 0&25 & 0&0095(3)\\
			S  & $4c$ & 0 & 0&69086(7) & 0&25 & 0&0099(2)\\
			O  & $4a$ & 0 & 0& & 0& & 0&01410(14)\\ \hline
		\end{tabular}
	\end{table}
	
	\begin{table}[htb]
		\caption{\label{tab:NXD100b}Refined atomic positions and magnetic moments for BaCoSO at 100\,K.  The moments $m_i$ along each axis $i$ are given in $\mu_B$.}
		\begin{tabular}{lccr@{.}lr@{.}lr@{.}lccr@{.}l}\\ \hline
			Site & Mult.& $x$ & \multicolumn{2}{c}{$y$} & \multicolumn{2}{c}{$z$} & \multicolumn{2}{c}{$U_{iso}$} & $m_x$ & $m_y$ & \multicolumn{2}{c}{$m_z$}\\ \hline \hline
			Ba & $4c$ & 0 & 0&38663(6) & 0&25 & 0&00285(13) & 0 & 0 & 0\\
			Co & $4c$ & 0 & 0&09522(11) & 0&25 & 0&0032(3) & 0 & 0 & 3&017(15)\\
			S  & $4c$ & 0 & 0&69136(12) & 0&25 & 0&0043(2) & 0 & 0 & 0\\
			O  & $4a$ & 0 & 0& & 0& & 0&00655(15) & 0 & 0 & 0\\ \hline
		\end{tabular}
	\end{table}
	
	\begin{table}[htb]
		\caption{\label{tab:NXD300c}Refined anisotropic displacement parameters in \AA$^2$\ for BaCoSO at room temperature.}
		\begin{tabular}{lr@{.}lr@{.}lr@{.}lccr@{.}l}\hline
			Site & \multicolumn{2}{c}{$U_{11}$} & \multicolumn{2}{c}{$U_{22}$} & \multicolumn{2}{c}{$U_{33}$} & $U_{12}$ & $U_{13}$ & \multicolumn{2}{c}{$U_{23}$} \\ \hline\hline
			Ba & 0&0080(2) & 0&0103(2) & 0&0083(3) & 0 & 0 & 0& \\
			Co & 0&0088(4) & 0&0094(3) & 0&0103(5) & 0 & 0 & 0& \\
			S  & 0&0078(4) & 0&0100(3) & 0&0119(5) & 0 & 0 & 0& \\
			O  & 0&0164(3) & 0&0138(2) & 0&0121(3) & 0 & 0 & $-$0&00497(15)\\ \hline
		\end{tabular}
	\end{table}
	
\end{document}

\title{Magnetic and Electronic Properties of Single-Crystalline BaCoSO:\\ Supplemental Material}
	
	\author{Yuanhe Song}
	\author{Xianshi Liu}
	\affiliation{State Key Laboratory of Surface Physics, Department of Physics, and Advanced Materials Laboratory, Fudan University, Shanghai 200438, China}
	
	\author{Dong Wu}
	\affiliation{International Center for Quantum Materials, School of Physics, Peking University, Beijing 100871, China}
	
	\author{Qi Yao}
	\author{Chenhaoping Wen}
	\affiliation{State Key Laboratory of Surface Physics, Department of Physics, and Advanced Materials Laboratory, Fudan University, Shanghai 200438, China}
	
	\author{Maxim Avdeev}
	\email{max@ansto.gov.au}
	\affiliation{Australian Nuclear Science and Technology Organisation, Lucas Heights, NSW 2234, Australia}
	
	\author{Rui Peng}
	\author{Haichao Xu}
	\author{Jin Miao}
	\author{Xia Lou}
	\author{Yifei Fang}
	\author{Binglin Pan}
	\affiliation{State Key Laboratory of Surface Physics, Department of Physics, and Advanced Materials Laboratory, Fudan University, Shanghai 200438, China}
	
	\author{Nanlin Wang}
	\affiliation{International Center for Quantum Materials, School of Physics, Peking University, Beijing 100871, China}
	\affiliation{Collaborative Innovation Center of Quantum Matter, Beijing 100871, China}
	
	\author{Darren C.\ Peets}
	\email{dpeets@nimte.ac.cn}
	\affiliation{State Key Laboratory of Surface Physics, Department of Physics, and Advanced Materials Laboratory, Fudan University, Shanghai 200438, China}
	\affiliation{Ningbo Institute of Materials Technology and Engineering, Chinese Academy of Sciences, Ningbo, Zhejiang 315201, China}
	
	\author{Donglai Feng}
	\email{dlfeng@fudan.edu.cn}
	\affiliation{State Key Laboratory of Surface Physics, Department of Physics, and Advanced Materials Laboratory, Fudan University, Shanghai 200438, China}
	\affiliation{Hefei National Laboratory for Physical Science at Microscale and Department of Physics, University of Science and Technology of China, Hefei, Anhui 230026, China}
	\affiliation{Collaborative Innovation Center of Advanced Microstructures, Nanjing 210093, China}
	
%	\date{\today}
	
	\maketitle
	
	\section{Charging effects and weak angle dependence in photoemission measurements}
	
	\begin{figure*}[htb]
		\includegraphics[width=\textwidth]{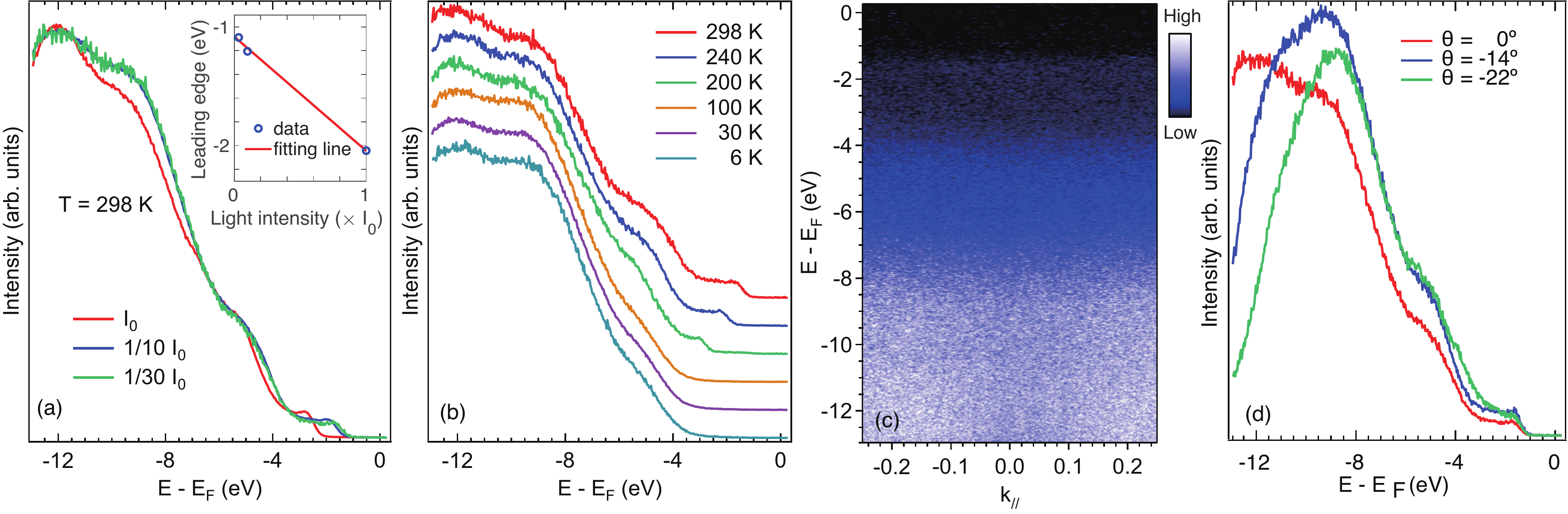}
		\caption{(a) Photoemission spectrum at 298\,K under different incident light intensities. The inset shows the relation between leading edge position and incident light intensity.  (b) Temperature dependence of the photoemission spectrum. (c) Angle-resolved spectrum at 298\,K. (d) Angle-integrated spectra centered at different polar emission angles at 298\,K. All spectra in (b-d) are taken under a light intensity of 1/30\,I$_0$.}
		\label{fig:charging}
	\end{figure*}
	
	We checked for charging problems in our photoemission measurements by investigating a BaCoSO sample under different photon fluxes at room temperature [Fig.~\ref{fig:charging}(a)]. Under the usual incident light intensity I$_0$, which provides $\sim$10$^{10}$\,cm$^{-2}$s$^{-1}$ effective photon flux, the spectrum shows a peak near $E_F$ at $-$2.83\,eV with a leading edge at $-$2.04\,eV. As the incident light intensity is decreased, peaks in the spectrum shift toward the Fermi level. In the spectrum with 1/30\,I$_0$, which has the least charging effect among the three, the position of the peak near $E_F$ is at $-$1.77\,eV and its leading edge is at $-$1.09\,eV. The shift between the spectra with 1/10 and 1/30\,I$_0$ is 0.12\,eV. By assuming a linear relation between the incident light intensities and the leading edge positions of the spectra, we estimate a lower limit for the band gap of BaCoSO in the limit of zero light intensity to be 1.08\,eV [see the inset to Fig.~\ref{fig:charging}(a)]. We also performed temperature-dependent measurements of the photoemission spectrum under a light intensity of 1/30\,I$_0$, and found that the peak close to $E_F$ was gradually suppressed on cooling and eventually vanished below 200\,K [Fig.~\ref{fig:charging}(b)]. This strong charging effect is consistent with the divergence at low temperature of an insulating resistivity. The angle-resolved spectrum of BaCoSO at room temperature [Fig.~\ref{fig:charging}(c)] shows no dispersive features, while the angle-integrated spectra centered at different polar emission angles indicate that there is some angle-dependence [Fig.~\ref{fig:charging}(d)].

%	\section{K-dosing-dependent ARPES at low temperature}
	
%	Fig.~\ref{fig:ARPES} presents additional results of our surface K-dosing ARPES study at 30\,K. Fig.~\ref{fig:ARPES}(a) shows the angle-resolved spectrum of BaCoSO within 13\,eV of the Fermi level ($E_F$) before K dosing, where no dispersing bands are observed. As a supplement to Fig.~7 in the main text, Fig.~\ref{fig:ARPES}(b) shows more precise photoemission measurements in a narrower energy range near $E_F$ [($-$0.25\,eV, +0.05\,eV)], which further demonstrate that any density of states is negligibly small and near the experimental noise level. This is reasonable since our temperature-dependence study unambiguously confirmed that BaCoSO has a severe charging problem below 200\,K, which suppresses the ARPES spectrum near $E_F$ at low temperature.

%	\begin{figure*}[htb]
%		\includegraphics[width=0.5\textwidth]{SM_ARPES.pdf}
%		\caption{ARPES study of surface K-dosed BaCoSO at 30\,K. (a) The angle-resolved spectrum of BaCoSO before K-dosing, showing no dispersing bands within 13\,eV of the Fermi level. The lack of a clear dispersion at high binding energies is common in insulators due to lifetime broadening. (b) A zoomed-in view of the angle-integrated spectra in Fig.~7 of the main text shows no DOS near $E_F$. The vertical dashed line marks $E_F$ while the horizontal dashed lines mark the zero intensity level for each spectrum.}
%		\label{fig:ARPES}
%	\end{figure*}
	
%	\bibliography{BaCoSO}